\documentclass{aip-cp}

\usepackage[numbers]{natbib}
\usepackage{rotating}
\usepackage{graphicx}
\usepackage{mathrsfs}
\usepackage{amssymb}
\usepackage{graphicx}
\usepackage[normalem]{ulem}
\usepackage[dvips]{color}
\usepackage{bm}
\usepackage{longtable}
\usepackage{times}

\renewcommand\sout{\bgroup \color{red} \ULdepth=-.5ex \ULset}

\renewcommand{\rm}[1]{\textrm{#1}}

\begin{document}

\title{Interface effects of strange quark matter}
\author[aff1]{Cheng-Jun~Xia}
\corresp[cor1]{Corresponding author and speaker: cjxia@itp.ac.cn}

\affil[aff1]{School of Information Science and Engineering, Ningbo Institute of Technology, Zhejiang University, Ningbo 315100, China}

\maketitle

\begin{abstract}
The interface effects play important roles for the properties of strange quark matter (SQM) and the related physical processes.
We show several examples on the implications of interface effects for both stable and unstable SQM. Based on an equivparticle
model and adopting mean-field approximation (MFA), the surface tension and curvature term of SQM can be obtained, which are
increasing monotonically with the density of SQM at zero external pressure. For a parameter
set constrained according to the 2$M_\odot$ strange star, we find the surface tension is $\sim$2.4 MeV/fm${}^2$, while it is
larger for other cases.
\end{abstract}

\section{\label{sec:intro}Introduction}

At ultra-high densities, QCD can be solved with perturbative approaches and suggests that quarks are no longer confined
within hadrons. In such cases, strange quark matter (SQM) is formed, which is comprised of approximately equal numbers of
$u$, $d$, and $s$ quarks. However, at physically accessible densities, perturbative approaches do not apply while lattice
QCD suffers the infamous sign problem. It is thus unclear about the density region where deconfinement phase transition
takes place, or the properties of SQM. And we have to rely on effective models due to the non-perturbative nature of SQM,
where many possibilities exist. For example, it was long suspected that SQM is more stable than nuclear matter, i.e., the
true ground state of strongly interacting system~\cite{Bodmer1971_PRD4-1601, Witten1984_PRD30-272, Terazaw1989_JPSJ58-3555}.
If true, there may exist stable lumps of SQM, e.g., strangelets~\cite{Farhi1984_PRD30-2379, Berger1987_PRC35-213,
Gilson1993_PRL71-332, Peng2006_PLB633-314}, nuclearites~\cite{Rujula1984_Nature312-734, Lowder1991_NPB24-177}, meteorlike
compact ultradense objects~\cite{Rafelski2013_PRL110-111102}, and strange stars~\cite{Itoh1970_PTP44-291, Alcock1986_ApJ310-261,
Haensel1986_AA160-121}. Nevertheless, in recent years it was realized that SQM may be unstable considering the dynamical
chiral symmetry breaking~\cite{Buballa1999_PLB457-261, Klahn2015_ApJ810-134}. Then only in extreme conditions it will
persists, e.g., in the centre of compact stars~\cite{Weber2005_PPNP54-193, Maruyama2007_PRD76-123015, Peng2008_PRC77-065807,
Shao2013_PRD87-096012, Klahn2013_PRD88-085001, Zhao2015_PRD92-054012, Li2015_PRC91-035803} or heavy-ion
collisions~\cite{Greiner1987_PRL58-1825, Greiner1991_PRD44-3517}.

\begin{figure}[h!]
\begin{minipage}[t]{0.49\linewidth}
\centering
\includegraphics[width=\textwidth]{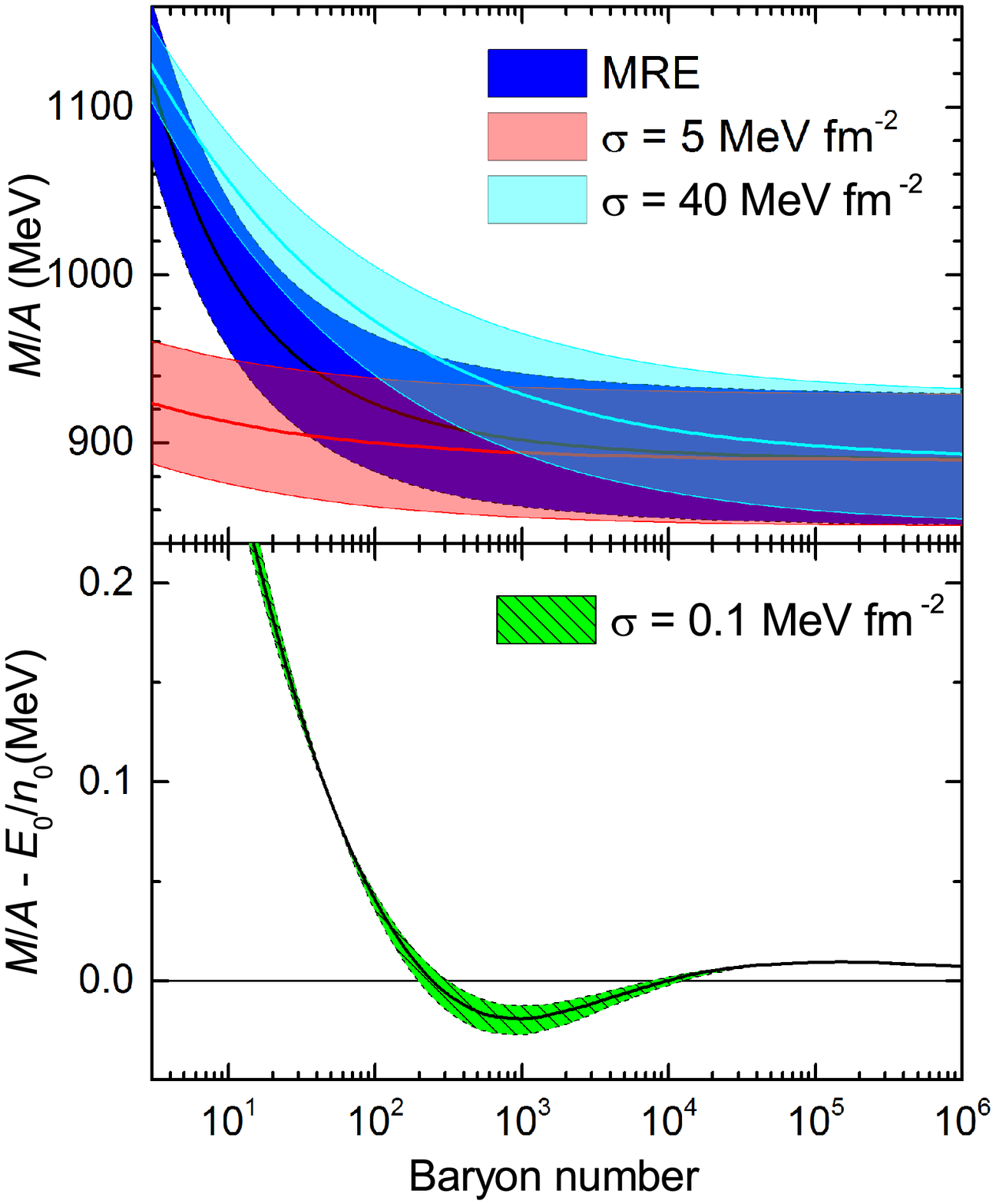}
\end{minipage}%
\hfill
\begin{minipage}[t]{0.49\linewidth}
\centering
\includegraphics[width=\textwidth]{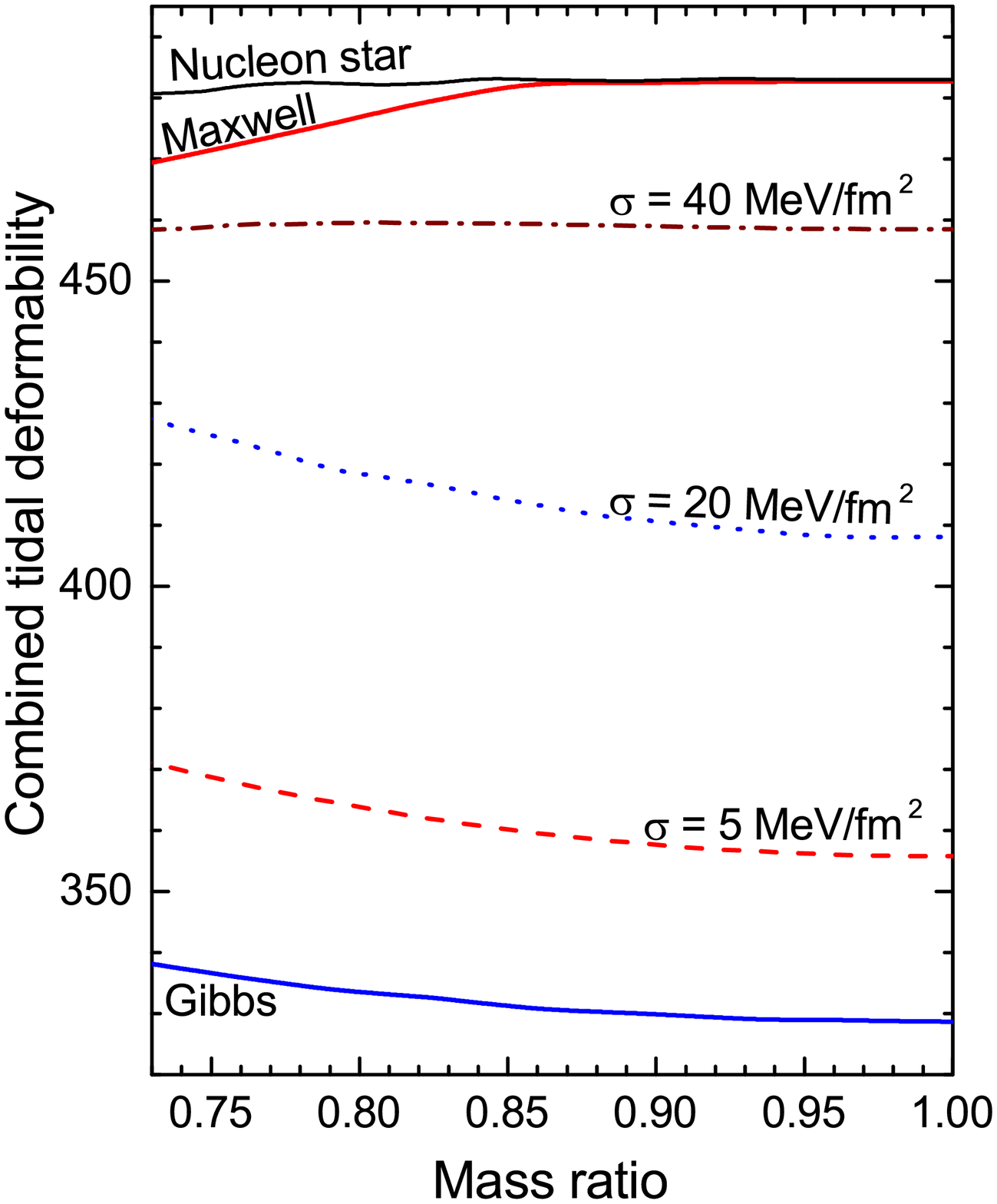}
\end{minipage}
\caption{ {Left:} \label{Fig:example} The energy per baryon ($M/A$) and energy excess per baryon $\left(M/A - E_0/n_0\right)$ of
strangelets obtained with various surface treatments, where the full lines and shaded regions correspond to the central
values and uncertainty of $B^{1/4} = 152\pm7$ MeV.
{Right:} The dimensionless combined tidal deformabilities of hybrid stars as functions of their mass ratio.
The data are taken from Refs.~\cite{Xia2016_PRD93-085025, Xia2019}, for more details please refer to these references.}
\end{figure}

For absolutely stable SQM, it is found that the properties of SQM objects are sensitive to the quark-vacuum
interface~\cite{Farhi1984_PRD30-2379, Berger1987_PRC35-213, Berger1989_PRD40-2128, Heiselberg1993_PRD48-1418}. To show this
explicitly, as an example, in the left panel of Fig.~\ref{Fig:example} we present the energy per baryon and energy excess
per baryon of strangelets obtained based on a unified description for SQM objects, i.e., the UDS model~\cite{Xia2016_SciBull61-172, Xia2016_SciSinPMA46-012021_E, Xia2016_PRD93-085025, Xia2017_JPCS861-012022, Xia2017_NPB916-669}, where both a constant surface
tension $\sigma$ and the multiple reflection
expansion (MRE) method~\cite{Berger1987_PRC35-213, Madsen1993_PRL70-391, Madsen1993_PRD47-5156, Madsen1994_PRD50-3328} are adopted.
The bag constant $B$ is fixed so that SQM is stable while two-flavor quark matter remains unstable with respect to nuclear
matter. In general, the energy per baryon is decreasing with the baryon number $A$. At small $A$, the interface effects are
important and destabilize a strangelet substantially~\cite{Farhi1984_PRD30-2379}. For the cases with ${M}/{A}>939$ MeV,
a strangelet will quickly decay into nucleons via neutron emission, which may be related to the kilonova of the possible binary
strange star merge events~\cite{Paulucci2017_IJMPCS45-1760042, Lai2018_RAA18-024}. Based on the derived mass formula of
strangelets, it was found that the minimum baryon number (at $M/A=939$ MeV) for metastable strangelets increases linearly
with $\sigma^3$~\cite{Berger1987_PRC35-213, Berger1989_PRD40-2128}. Meanwhile, as indicated in the upper-left of
Fig.~\ref{Fig:example}, the curvature contribution treated with the MRE method~\cite{Madsen1993_PRL70-391, Madsen1993_PRD47-5156,
Madsen1994_PRD50-3328} is also important and further destabilize small strangelets. Considering the effects
of charge screening~\cite{Heiselberg1993_PRD48-1418}, as indicated in the lower-left of Fig.~\ref{Fig:example},
adopting a small enough $\sigma$ would predict strangelets ($A\approx 1000$) that are more stable than
others~\cite{Alford2006_PRD73-114016}. In such cases, larger strangelets will go fission, and the surfaces of strange stars
will fragment into crystalline crusts consist of strangelets and electrons~\cite{Jaikumar2006_PRL96-041101}, which may even
form low-mass large-radius strangelet dwarfs~\cite{Alford2012_JPG39-065201}. For larger SQM objects, the interface effects
have little impact on their masses. However, the charge properties are greatly affected by the interface effects, e.g.,
the electron-positron pair creation on the surface affects significantly the maximum net charge an object can
carry~\cite{Madsen2008_PRL100-151102}, and the effects of quark depletion and charge screening~\cite{Xia2017_JPCS861-012022}
also play important roles on the surface charge properties of large SQM objects.

In the cases of unstable SQM, at the centre of hybrid stars, it may coexists with hadronic matter. Due to the relocation of
charged particles on the quark-hadron interface, the geometrical structures such as droplet, rod, slab, tube, and bubble
start to emerge~\cite{Heiselberg1993_PRL70-1355, Voskresensky2002_PLB541-93, Tatsumi2003_NPA718-359, Voskresensky2003_NPA723-291,
Endo2005_NPA749-333, Maruyama2007_PRD76-123015, Yasutake2012_PRD86-101302}. Nevertheless, it was found that these structures
are affected significantly by the quark-hadron interface~\cite{Maruyama2007_PRD76-123015}. In particular, the sizes of the
geometrical structures increase with the quark-hadron interface tension $\sigma$. When $\sigma$ is greater than a critical
value $\sigma_\mathrm{c}$, the Maxwell construction is effectively restored, i.e., bulk separation of quark and hadron phases.
Meanwhile, for a vanishing $\sigma$, the geometrical structures become small enough and the corresponding quark-hadron mixed
phase approaches to the cases obtained with the Gibbs construction. The structural differences caused by introducing different
$\sigma$ could affect many physical processes in hybrid stars~\cite{Reddy2000_PLB475-1, Maruyama2008_PLB659-192,
Prakash1992_ApJ390-L77, Tatsumi2003_PTP110-179, Takatsuka2006_PTP115-355, Ayriyan2018_PRC97-045802}. In the right panel of
Fig.~\ref{Fig:example} we show an example of the interface effects on the dimensionless combined tidal deformabilities
$\tilde{\Lambda}$ of hybrid stars in light of the observed binary neutron star merger event GW170817~\cite{LVC2019_PRX9-011001}.
It is found that varying $\sigma$ has sizable effects on $\tilde{\Lambda}$, while similar cases are expected for the radii of
hybrid stars~\cite{Xia2019}.

For the dynamic processes where the transition between SQM and hadronic matter takes place, the interface effects
also play important roles. For example, it was shown that a larger $\sigma$ inhibits quark matter nucleation during
the deconfinement phase transitions in cold neutron stars~\cite{Bombaci2016_EPJA52-58, Lugones2016_EPJA52-53}, core-collapse
supernova~\cite{Mintz2010_PRD81-123012, Mintz2010_JPG37-094066}, and heavy-ion collisions~\cite{Toro2006_NPA775_102-126,
Fogaca2016_PRC93-055204}. Similarly, the interface effects are important for the hadronization phase transition in the early
Universe~\cite{Alcock1985_PRD32-1273, Alcock1989_PRD39-1233, Lugones2004_PRD69-063509, Li2015_AP62-115}.

Due to the crucial significance, it is essential that we have a full understanding on the interface effects of SQM.
Based on lattice QCD, the surface tension can be estimated for vanishing chemical potentials~\cite{Huang1990_PRD42-2864,
Huang1991_PRD43-2056, Alves1992_PRD46-3678, Brower1992_PRD46-2703, Forcrand2005_NPB140-647, Forcrand2005_PRD72-114501}.
However, such type of calculations break down for finite chemical potentials due to the infamous sign problem,
then we have to rely on effective models. The recent estimations for the surface tension indicate $\sigma= 5 \sim 30\
\mathrm{MeV/fm}^{2}$ according to the MIT bag model~\cite{Oertel2008_PRD77-074015}, linear
sigma model~\cite{Palhares2010_PRD82-125018, Pinto2012_PRC86-025203, Kroff2015_PRD91-025017}, Nambu-Jona-Lasinio
model~\cite{Garcia2013_PRC88-025207, Ke2014_PRD89-074041}, three-flavor Polyakov-quark-meson model~\cite{Mintz2013_PRD87-036004},
and Dyson-Schwinger equation approach~\cite{Gao2016_PRD94-094030}. Adopting the MRE method, lager values of $\sigma$
were obtained based on the quasiparticle model with $\sigma= 30 \sim 70\ \mathrm{MeV/fm}^{2}$~\cite{Wen2010_PRC82-025809}
and Nambu-Jona-Lasinio model with $\sigma= 145 \sim 165\ \mathrm{MeV/fm}^{2}$~\cite{Lugones2013_PRC88-045803,
Lugones2017_PRC95-015804}. A largest value was obtained in early estimations for color-flavor locked SQM, i.e.,
$\sigma\approx 300\ \mathrm{MeV/fm}^{2}$~\cite{Alford2001_PRD64-074017}.

\section{\label{sec:the_Lagrangian} Equivparticle model for strangelets}

In this work, as was done in Ref.~\cite{Xia2018_PRD98-034031}, we adopt the equivparticle model~\cite{Peng2000_PRC62-025801,
Wen2005_PRC72-015204, Wen2007_JPG34-1697, Xia2014_SCPMA57-1304, Chen2012_CPC36-947, Chang2013_SCPMA56-1730, Xia2014_PRD89-105027,
Chu2014_ApJ780-135, Hou2015_CPC39-015101, Peng2016_NST27-98, Chu2017_PRD96-083019} to study the interface effects of SQM.
All the strong interactions are treated with density-dependent quark masses in the equivparticle model, while the quarks are
considered to be quasi-free particles. Here the following quark mass scaling is adopted
\begin{equation}
  m_i(n_{\mathrm b})=m_{i0} + m_\mathrm{I}(n_{\mathrm b})=m_{i0}+\frac{D}{\sqrt[3]{n_\mathrm{b}}}+C\sqrt[3]{n_\mathrm{b}},  \label{Eq:mnbC}
\end{equation}
where $m_{i0}$ is the current mass of quark flavor $i$~\cite{PDG2016_CPC40-100001}, $m_\mathrm{I}$ arises due to the strong
interaction between quarks, and $n_\mathrm{b}= \sum_{i=u,d,s}n_i/3$ is the baryon number density with the number density
$n_i=\langle \bar{\Psi}_i \gamma^0 \Psi_i\rangle$. Note that the inversely cubic scaling in Eq.~(\ref{Eq:mnbC}) corresponds
to the linear confinement interaction, while the cubic scaling represents the one-gluon-exchange interaction for
$C<0$~\cite{Chen2012_CPC36-947} and the leading-order perturbative interaction for $C>0$~\cite{Xia2014_PRD89-105027}.
The Lagrangian density is then given by
\begin{equation}
\mathcal{L} =  \sum_{i=u,d,s} \bar{\Psi}_i \left[ i \gamma^\mu \partial_\mu - m_i(n_\mathrm{b}) - e q_i \gamma^\mu A_\mu \right]\Psi_i
             - \frac{1}{4} A_{\mu\nu}A^{\mu\nu},  \label{eq:Lgrg_all}
\end{equation}
where $\Psi_i$ is the Dirac spinor, $m_i(n_\mathrm{b})$ the equivalent quark mass obtained with Eq.~(\ref{Eq:mnbC}), and
$A_\mu$ the photon field with the field tensor $A_{\mu\nu} = \partial_\mu A_\nu - \partial_\nu A_\mu$. With the mean-field
and no-sea approximations, and assuming spherical symmetry for a strangelet, applying a standard variational procedure
on the Lagrangian density~(\ref{eq:Lgrg_all}) gives the Klein-Gordon equation for photons
\begin{equation}
- \nabla^2 A_0 = e n_\mathrm{ch} \label{Eq:K-G}
\end{equation}
and the Dirac equation for quarks
\begin{equation}
 \left(\begin{array}{cc}
  V_i + V_S  + m_{i0}                                         & {\displaystyle -\frac{\mbox{d}}{\mbox{d}r} + \frac{\kappa}{r}}\\
  {\displaystyle \frac{\mbox{d}}{\mbox{d}r}+\frac{\kappa}{r}} & V_i - V_S - m_{i0}                      \\
 \end{array}\right)
 \left(\begin{array}{c}
  G_{n\kappa} \\
  F_{n\kappa} \\
 \end{array}\right)
 = \varepsilon_{n\kappa}
 \left(\begin{array}{c}
  G_{n\kappa} \\
  F_{n\kappa} \\
 \end{array}\right) \:
\label{Eq:RDirac}
\end{equation}
with the Dirac spinor expanded by
\begin{equation}
 \psi_{n\kappa m}({\bm r}) =\frac{1}{r}
 \left(\begin{array}{c}
   iG_{n\kappa}(r) \\
    F_{n\kappa}(r) {\bm\sigma}\cdot{\hat{\bm r}} \\
 \end{array}\right) Y_{jm}^l(\theta,\phi)\:.
\label{EQ:RWF}
\end{equation}
Here $G_{n\kappa}(r)/r$ and $F_{n\kappa}(r)/r$ are the radial wave functions and $Y_{jm}^l(\theta,\phi)$ the spinor spherical
harmonics with the quantum number $\kappa$ related to the angular momenta $(l,j)$ with $\kappa=(-1)^{j+l+1/2}(j+1/2)$.
The charge density in Eq.~(\ref{Eq:K-G}) is given by $n_\mathrm{ch} = \sum_i q_i n_i$ with $q_u = 2/3, q_d = -1/3$, and $q_s = -1/3$.
In  Eq.~(\ref{Eq:RDirac}), $\varepsilon_{n\kappa}$ is the single particle energy, while the mean field scalar and vector potentials
are obtained with
\begin{eqnarray}
 V_S &=& m_\mathrm{I}(n_\mathrm{b}), \label{Eq:Vs}\\
 V_i &=& \frac{1}{3}\frac{\mbox{d} m_\mathrm{I}}{\mbox{d} n_\mathrm{b}}\sum_{i=u,d,s}  n_i^\mathrm{s} + e q_i A_0. \label{Eq:Vv}
\end{eqnarray}
Note that the vector potentials in Eq.~(\ref{Eq:Vv}) share a common term $V_V = \frac{1}{3}\frac{\mbox{d} m_\mathrm{I}}{\mbox{d} n_\mathrm{b}}
\sum_{i=u,d,s}  n_i^\mathrm{s}$ for different types of quarks, which arises due to the density dependence of quark masses~\cite{Xia2018_PRD98-034031}.
In principle, for any model with density dependent masses or coupling constants, one needs to be cautious not to violate the self-consistency
of thermodynamics~\cite{Brown1991_PRL66-2720, Lenske1995_PLB345-355, Wang2000_PRC62-015204, Peng2000_PRC62-025801, Torres2013_EPL101-42003,
Dexheimer2013_EPJC73-2569, Xia2014_PRD89-105027}. Then the radial wave functions can be obtained by solving the Dirac equation~(\ref{Eq:RDirac}),
which gives the scalar and vector densities for quarks via
\begin{eqnarray}
 n_i^\mathrm{s}(r) &=& \frac{1}{4\pi r^2}\sum_{k=1}^{N_i} \left[|G_{k i}(r)|^2-|F_{k i}(r)|^2\right], \label{Eq:ns} \\
 n_i(r) &=& \frac{1}{4\pi r^2}\sum_{k=1}^{N_i} \left[|G_{k i}(r)|^2+|F_{k i}(r)|^2\right]. \label{Eq:nv}
\end{eqnarray}

At given $C$ and $D$, the differential equations~(\ref{Eq:K-G}) and (\ref{Eq:RDirac}) for a strangelet with given baryon number $A$ are solved
in an iterative manner:
\begin{enumerate}
  \item \label{item:itr_1} Assuming initial scalar and vector densities $n_i^\mathrm{s}(r)$ and $n_i(r)$;
  \item \label{item:itr_2} Obtain the mean field potentials based on Eqs.~(\ref{Eq:Vs}) and (\ref{Eq:Vv}), where the Coulomb potential $A_0(r)$ is determined by solving the Klein-Gordon equation~(\ref{Eq:K-G});
  \item \label{item:itr_3} The radial wave functions are fixed by solving the Dirac equation~(\ref{Eq:RDirac});
  \item \label{item:itr_4} Fill quarks in the levels corresponding to the lowest single particle energies $\varepsilon_{n\kappa}$;
  \item \label{item:itr_5} With the scalar and vector densities determined by Eqs.~(\ref{Eq:ns}) and (\ref{Eq:nv}), go to step~\ref{item:itr_2} until
  convergence is reached.
\end{enumerate}
Finally, the total quark number $N_i\ (i=u,d,s)$ and mass of a strangelet in $\beta$-equilibrium can be obtained with
\begin{eqnarray}
   N_i &=& \int 4\pi r^2 n_i(r) \mbox{d}r, \label{Eq:axi} \\
   M &=& \sum_{i=u,d,s}\sum_{k=1}^{N_i}\varepsilon_{ki} - \int 12\pi r^2 n_\mathrm{b}(r) V_V(r) \mbox{d}r  - \int 2\pi r^2 n_\mathrm{ch}(r) e A_0(r) \mbox{d}r. \label{Eq:M}
\end{eqnarray}
Note that the iteration runs inside a box in coordinate space with the grid width $0.005$ fm, while the box size $R$ is fixed at vanishing densities.

\section{\label{sec:surf} Interface effects of SQM}

\begin{figure}[h!]
\centering
\includegraphics[width=10cm]{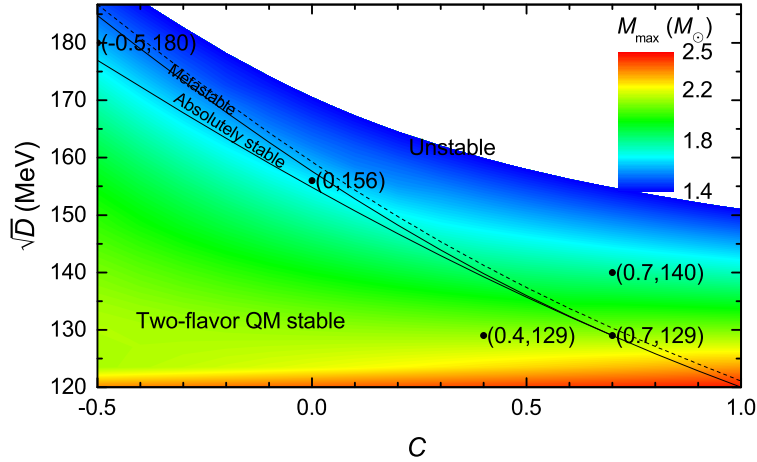}
\caption{\label{Fig:MmaxCD} The maximum mass $M_\mathrm{max}$ of strange stars as a function of the parameter $C$ and $\sqrt{D}$,
while the stability window of SQM is indicated with the 3 black curves. The data are taken from Refs.~\cite{Xia2014_PRD89-105027,
Xia2015_CAA56-79, Xia2016_NSC2015, Peng2018_JPSCP20-011022}, for more details please refer to these references.  }
\end{figure}

To fix the parameters $C$ and $\sqrt{D}$ for the quark mass scaling in Eq.~(\ref{Eq:mnbC}), in Fig.~\ref{Fig:MmaxCD} we present
the maximum mass $M_\mathrm{max}$ of strange stars in the parameter space. The 3 black curves divide the whole area into 4 regions.
Above the dashed curve, SQM is unstable, while below the solid curve is the forbidden region where two-flavor quark matter is stable.
For the regions between the curves SQM is absolutely stable (solid-solid region) and metastable (solid-dashed region). It is found
that a strange star can be more massive than PSR J0348+0432~\cite{Antoniadis2013_Science340-1233232} if we take $C\gtrsim 0.6$ in the
solid-solid region~\cite{Xia2014_PRD89-105027, Xia2015_CAA56-79, Xia2016_NSC2015, Peng2018_JPSCP20-011022}. For unstable SQM, according
to Ref.~\cite{Li2015_PRC91-035803}, a hybrid star can be more massive than 2$M_\odot$ if we adopt $C = 0.7$ and $\sqrt{D} = 170$, 190 MeV.
To consider all possibilities, as indicated in Fig.~\ref{Fig:MmaxCD} with the black dots, we adopt the parameter sets ($C$, $\sqrt{D}$):
($-0.5$, 180 MeV), (0, 156 MeV), (0.4, 129 MeV), (0.7, 129 MeV), and (0.7, 140 MeV).

\begin{figure}[h!]
\begin{minipage}[t]{0.49\linewidth}
\centering
\includegraphics[width=\textwidth]{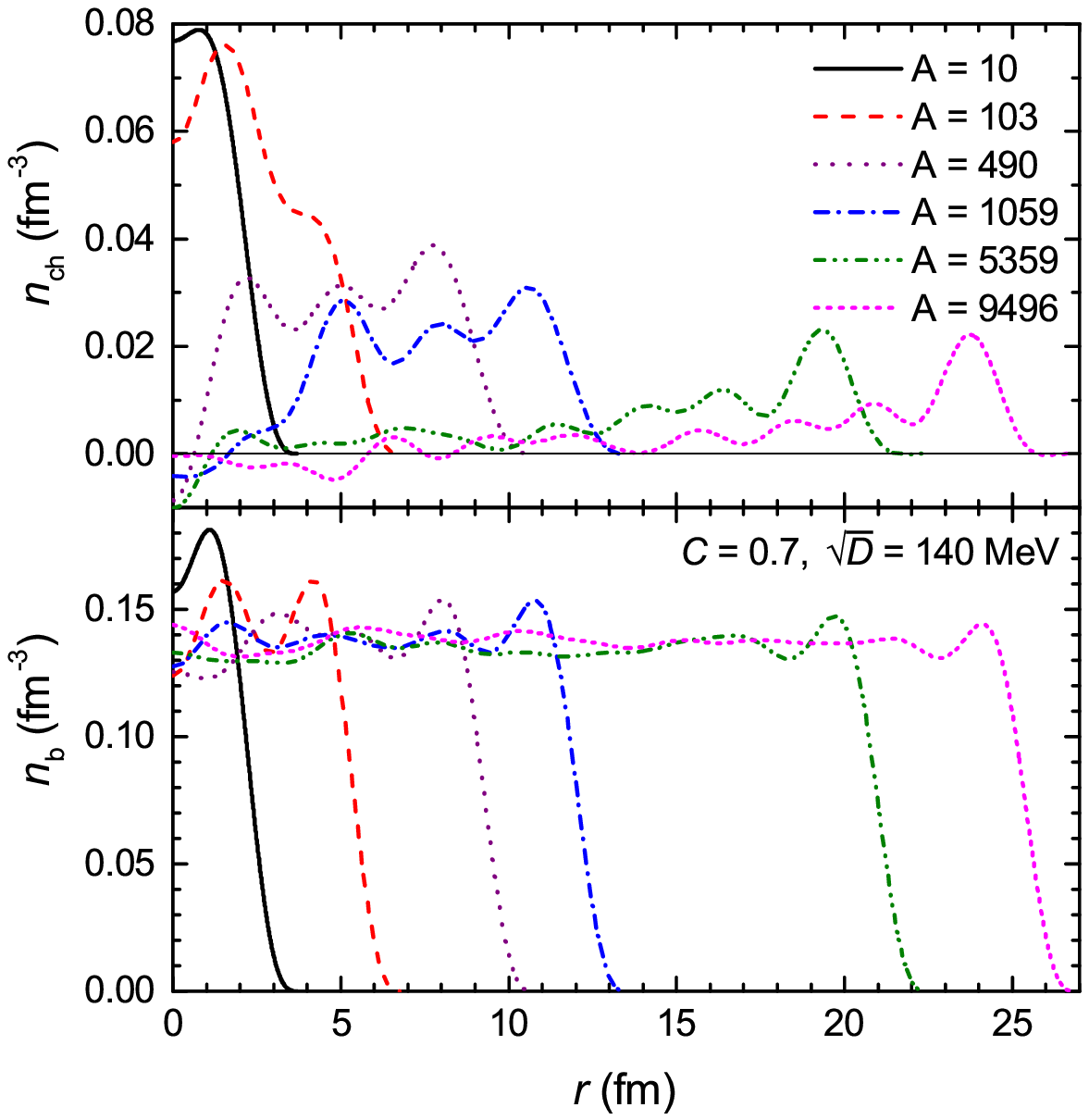}
\end{minipage}%
\hfill
\begin{minipage}[t]{0.475\linewidth}
\centering
\includegraphics[width=\textwidth]{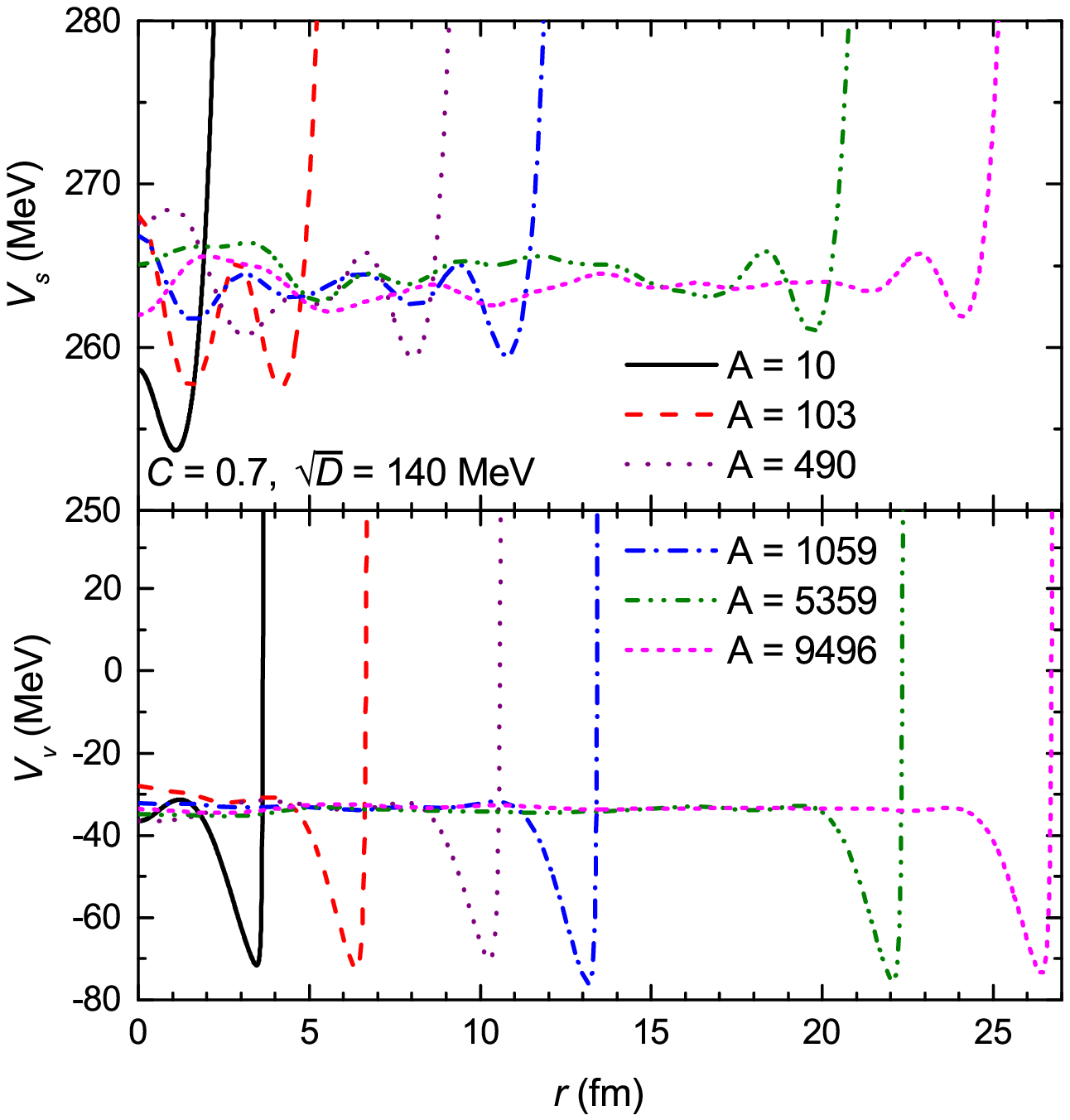}
\end{minipage}
\caption{ {Left:} \label{Fig:nbchV_C07D140} Baryon and charge densities inside strangelets with various baryon numbers.
{Right:} The obtained scalar and vector potentials.}
\end{figure}

With the selected parameter sets, we apply iteration procedure with the steps~\ref{item:itr_1}-\ref{item:itr_5} and obtain
the properties of strangelets. As an example, in left panel of Fig.~\ref{Fig:nbchV_C07D140} we present the baryon number
density and charge density for strangelets obtained with various baryon numbers $A$, where the parameter set $C = 0.7$ and
$\sqrt{D} = 140$ MeV is adopted. It is found that the size of a strangelet increases with $A$ while the internal baryon
number density becomes smooth and approaches to the bulk value at $n_0 = 0.13\ \mathrm{fm}^{-3}$ as indicated in
Table~\ref{table:prop}. The density starts to drop at $R-r\approx 2$-3 fm and reaches zero at $r=R$, which forms the
surface of a strangelet. Note that the density profiles on the surface varies slightly with $A$ and starts to converge
at $A\gtrsim 10^5$, where the variations can be attributed to the curvature term~\cite{Xia2018_PRD98-034031}. Due to
Coulomb repulsion, internally the charge density is decreasing with $A$, while the surface charge density persists and
starts to converge at large enough baryon numbers. This is consistent with our findings with the UDS model, where
a constant surface charge density was obtained for large enough SQM objects~\cite{Xia2016_SciBull61-172,
Xia2016_SciSinPMA46-012021_E, Xia2016_PRD93-085025, Xia2017_JPCS861-012022, Xia2017_NPB916-669}.

The mean field potentials can then be obtained with the density profiles based on Eq.~(\ref{Eq:Vs}) and Eq.~(\ref{Eq:Vv}).
The corresponding scalar and vector potentials are presented in the right panel of Fig.~\ref{Fig:nbchV_C07D140}.
Due to the inversely cubic scaling in Eq.~(\ref{Eq:mnbC}), the obtained potentials approaches to infinity at the
quark-vacuum interface, i.e., the linear confinement of quarks. Internally, it is found that the potential depth of
$V_S$ increases with $A$ while $V_V$ varies little. Similar to the cases on the density distributions of strangelets,
the mean field potentials in the vicinity of quark-vacuum interfaces vary slightly with $A$ and start to converge
at $A\gtrsim 10^5$.

\begin{table}[h!]
\caption{\label{table:prop} The bulk properties of SQM obtained at zero external pressure. The fitted liquid-drop
parameters and the corresponding surface tension and curvature term are given as well.}
\begin{tabular}{cc|ccccc|cccc} \hline \hline
\multicolumn{2}{c|}{Parameters} & \multicolumn{5}{c|}{Bulk properties}
& \multicolumn{4}{c}{Interface effects}   \\ \hline
$C$ & $\sqrt{D}$ & $n_0$ & $n_{u0}$ & $n_{d0}$ & $n_{s0}$ & ${E_0}/{n_0}$ & $\alpha_S$ & $\alpha_C$   & $\sigma$   &$\lambda$ \\
 & MeV & fm${}^{-3}$ & fm${}^{-3}$ & fm${}^{-3}$ & fm${}^{-3}$ & MeV & MeV & MeV & MeV/fm${}^2$ & MeV/fm  \\ \hline
-0.5& 180 & 0.37  & 0.37  & 0.49 & 0.26  & 900.04  & 58 & 328 & 6.3 & 15.2  \\
0   & 156 & 0.24  & 0.24  & 0.36 & 0.12  & 911.87  & 69 & 243 & 5.5 & 9.70  \\
0.4 & 129 & 0.11  & 0.11  & 0.20 & 0.023 & 850.91  & 56 & 177 & 2.7 & 5.49  \\
0.7 & 129 & 0.099 & 0.099 & 0.19 & 0.0055& 918.94  & 54 & 172 & 2.4 & 5.12  \\
0.7 & 140 & 0.13  & 0.13  & 0.24 & 0.018 & 995.77  & 61 & 185 & 3.3 & 6.03  \\
\hline
\end{tabular}
\end{table}

\begin{figure}[h!]
\begin{minipage}[t]{0.49\linewidth}
\centering
\includegraphics[width=\textwidth]{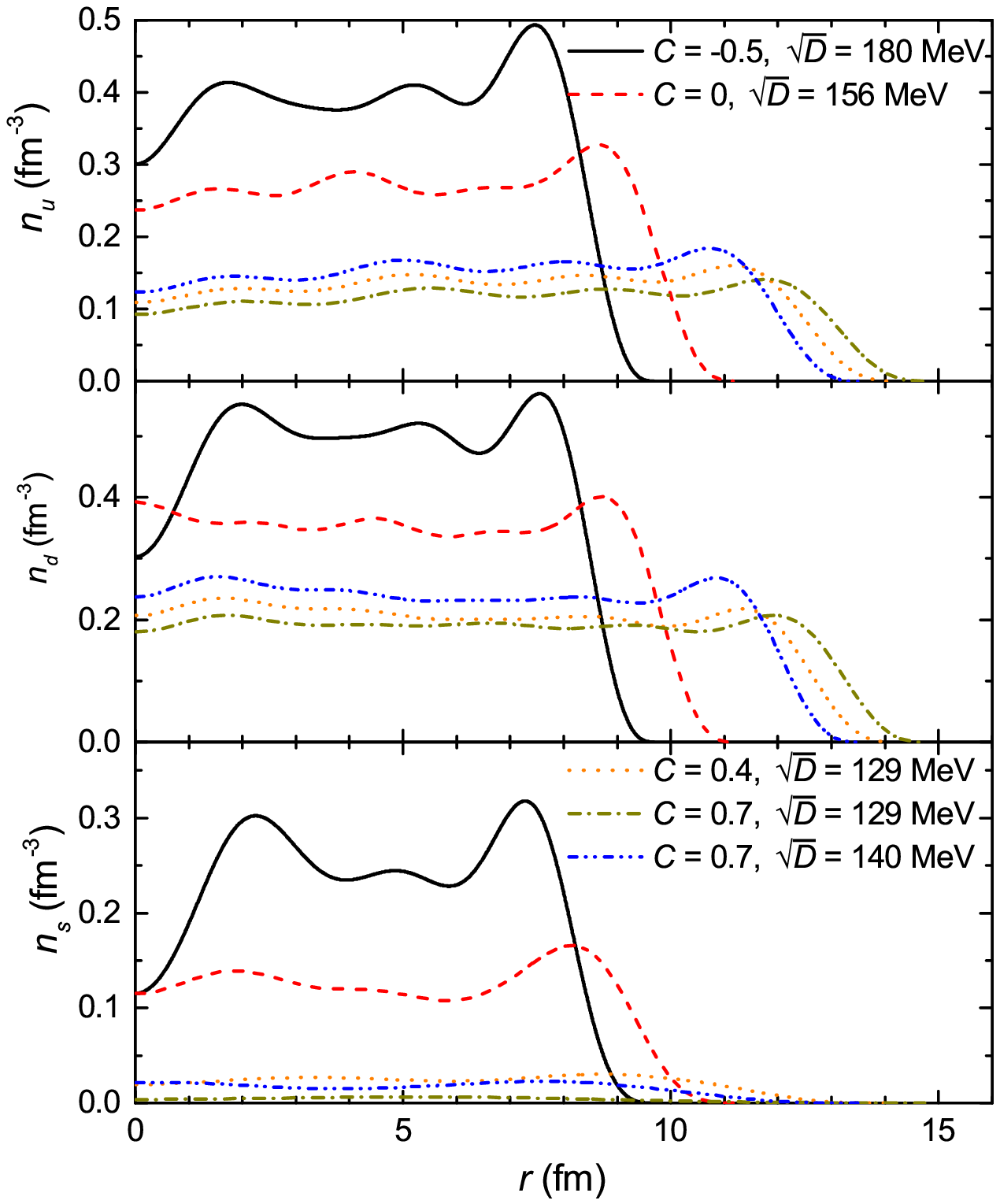}
\end{minipage}%
\hfill
\begin{minipage}[t]{0.49\linewidth}
\centering
\includegraphics[width=\textwidth]{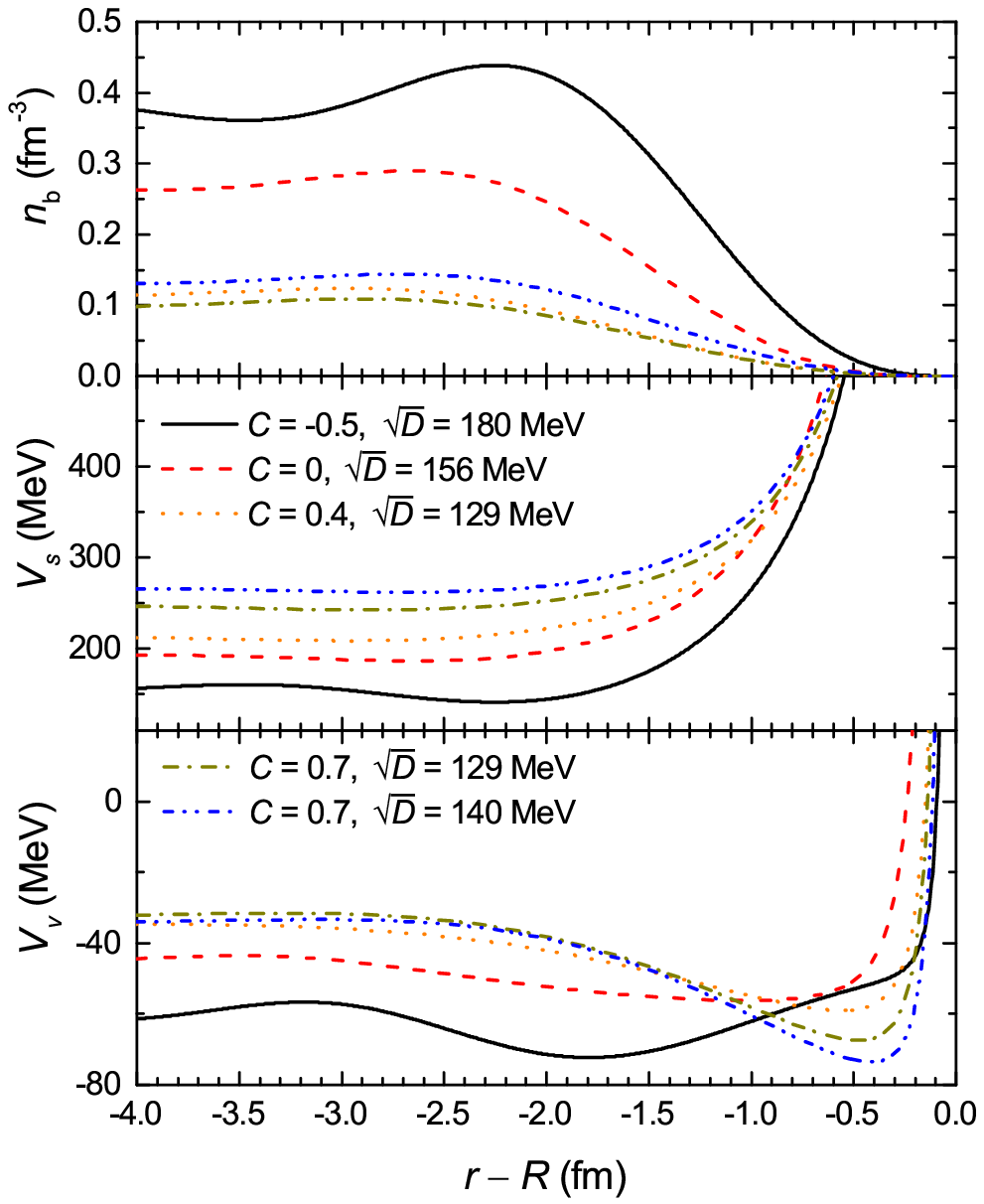}
\end{minipage}
\caption{ {Left:} \label{Fig:Surf_ni} Density profiles of $u$-, $d$-, $s$-quarks for strangelets with $A=1059$.
{Right:} The surface density and potential profiles for strangelets with $A\approx 10^5$.}
\end{figure}

To show the parameter dependence of strangelets' density profiles, in the left panel of Fig.~\ref{Fig:Surf_ni} we present
the density distributions of $u$-, $d$-, $s$-quarks for strangelets at $A=1059$. In general, the internal densities for each
species approach to their bulk values as indicated in Table~\ref{table:prop}, where a strangelet becomes more compact for larger
$\sqrt{D}$ and smaller $C$. This is consistent with our previous findings~\cite{Zhou2018_IJMPE27-1850037}. Due to the
large current mass, a strangelet carries less $s$-quarks than $u$-, $d$-quarks, then internally the number density for
$s$-quarks is smaller than those of $u$-, $d$-quarks. This becomes more evident when we adopt larger $C$, where $s$-quarks
are more diffused on strangelets' surfaces.

Since the curvature term becomes insignificant and the surface profiles converge for strangelets with
$A\gtrsim 10^5$~\cite{Xia2018_PRD98-034031}, the surface structures of strangelets ($A\approx 10^5$) are presented
in the right panel of Fig.~\ref{Fig:Surf_ni}  with the baryon number density, scalar and vector potentials in the
vicinity of quark-vacuum interfaces. Due to the linear confinement adopted with our mass scaling in Eq.~(\ref{Eq:mnbC}),
at $R-r\approx 2$-3 fm the obtained density starts to drop and slowly approaches to zero on the surface at $r=R$.
This is essentially different from the predictions of MIT bag model, where the density drops suddenly on the
surface~\cite{Oertel2008_PRD77-074015}. Since the lattice calculation predicts a linear confinement for
quarks~\cite{Belyaev1984_PLB136-273}, we believe our calculation is more reasonable. Since a larger $D$ results in
stronger confinement while increasing $C$ produces larger repulsive interaction, the internal density of a strangelet
increases with $D$ and decreases with $C$, which is in accordance with
the bulk density $n_0$ of SQM as indicated in Table~\ref{table:prop}. For larger $n_0$, it is found that the density
drops faster on the surface. Accordingly, the mean field potentials vary more drastically with $r$ on the surface,
which is expected to have a strong impact for the interface effects.

\begin{figure}[h!]
\begin{minipage}[t]{0.45\linewidth}
\centering
\includegraphics[width=\textwidth]{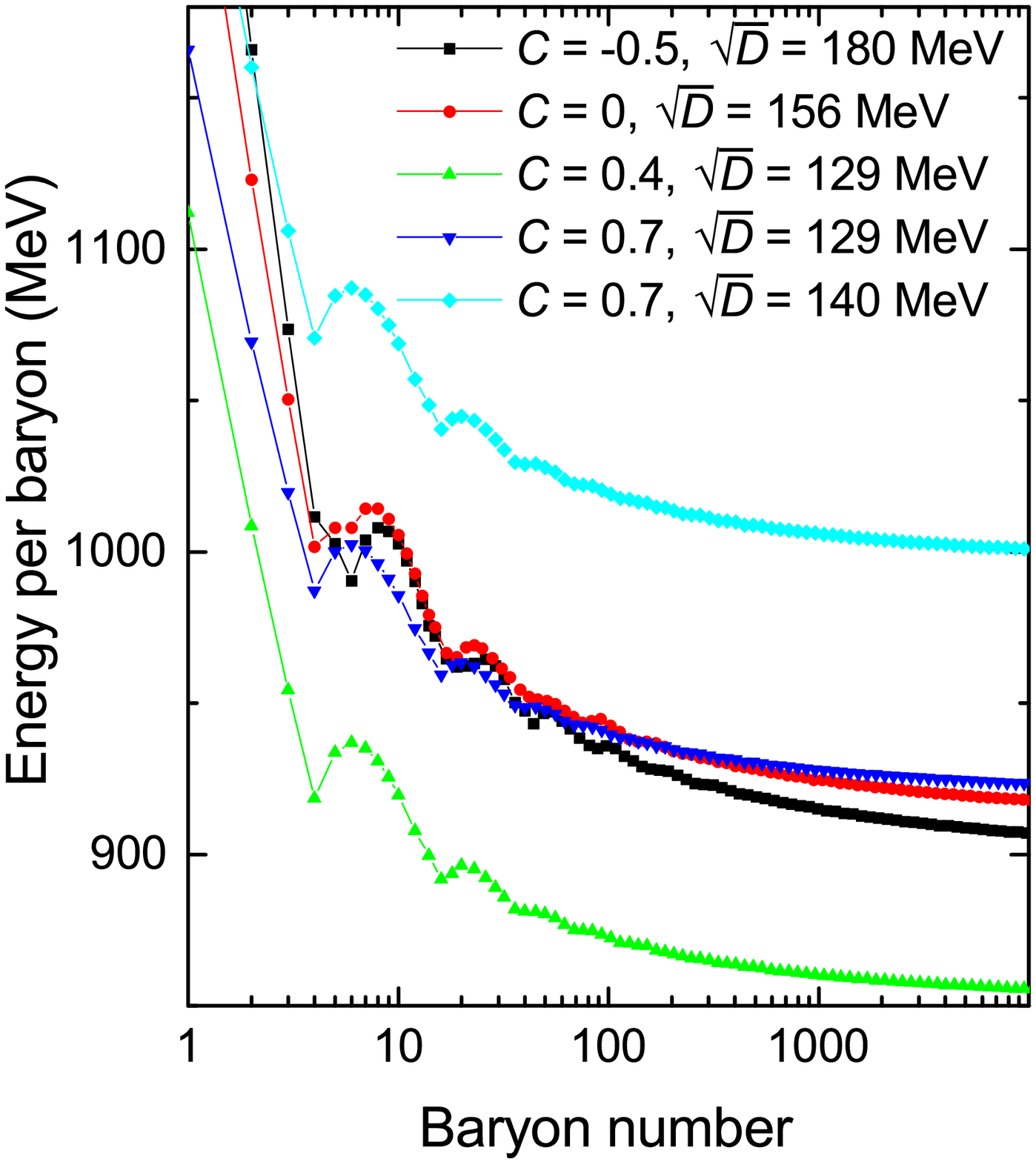}
\end{minipage}%
\hfill
\begin{minipage}[t]{0.4\linewidth}
\centering
\includegraphics[width=\textwidth]{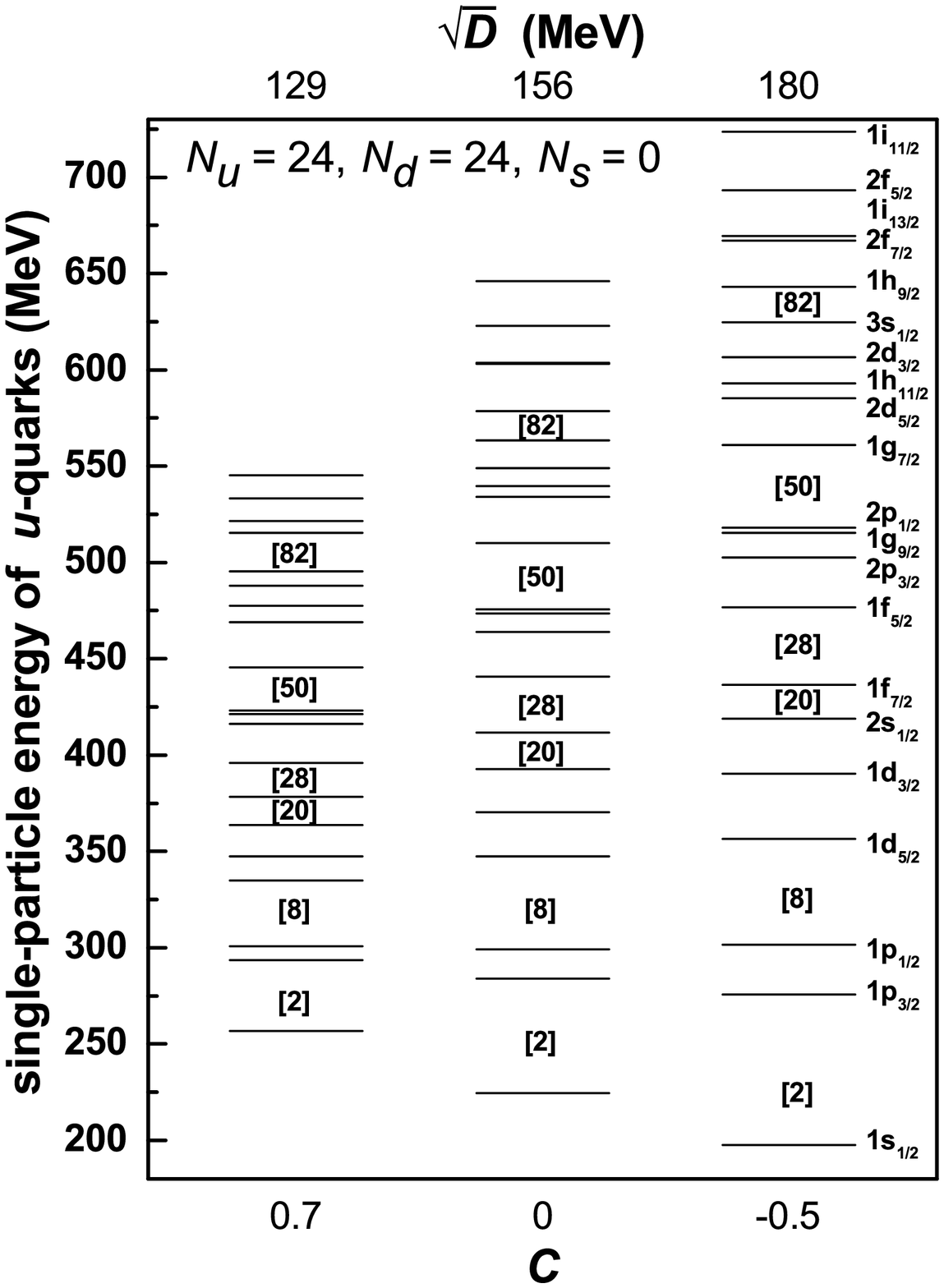}
\end{minipage}
\caption{ {Left:} \label{Fig:EpA_spl} The energy per baryon of strangelets in MFA.
{Right:} The single-particle levels for $u$-quarks. The magic numbers for finite nuclei are indicated with
the baryon numbers inside the square brackets.}
\end{figure}

In the left panel of Fig.~\ref{Fig:EpA_spl} we present the obtained energy per baryon of strangelets, which is decreasing with
$A$ and approaches to the bulk value in Table~\ref{table:prop}. Due to the shell effects, it is found that strangelets at $A=4$,
16, $\ldots$ are more stable than others. Meanwhile, if we adopt larger $\sqrt{D}$ and smaller $C$, strangelets at $A=6$, 18,
$\ldots$ becomes more stable since $s$-quarks appear, e.g., for the parameter sets ($C$, $\sqrt{D}$): ($-0.5$, 180 MeV) and
(0, 156 MeV). As indicated in the right panel of Fig.~\ref{Fig:EpA_spl}, those baryon numbers correspond to the magic numbers
6, 24, \ldots for $u$, $d$, $s$ quarks. The corresponding single-particle levels for quarks in strangelets are similar as
nucleons in ordinary nuclei~\cite{Sun2018_CPC42-25101}, predicting similar magic number shell effects at small baryon
numbers as indicated with the baryon numbers inside the square brackets in the right panel of Fig.~\ref{Fig:EpA_spl}.
For the cases with $A>16$, the magic number 20 does not appear and the magic number 28 appears only for certain choices of
parameter sets, while new magic numbers appear after 28, such as 34, 58, etc.

\begin{figure}[h!]
\centering
\includegraphics[width=9cm]{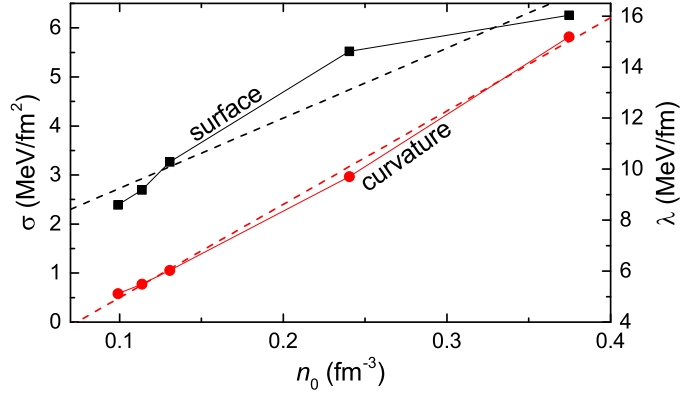}
\caption{\label{Fig:SigmaL} The obtained surface tension $\sigma$ and curvature term $\lambda$ of SQM as functions of the saturation
density $n_0$.}
\end{figure}

Based on the energy per baryon indicated in Fig.~\ref{Fig:EpA_spl}, the surface tension $\sigma$ and curvature term $\lambda$
of SQM can be obtained by fitting ${M}/{A}$ to a liquid-drop type formula~\cite{Oertel2008_PRD77-074015}
\begin{equation}
\frac{M}{A} = \frac{E_0}{n_0} + \frac{\alpha_S}{A^{1/3}} + \frac{\alpha_C}{A^{2/3}} \label{Eq:M_ld}
\end{equation}
with
\begin{equation}
\sigma  = \alpha_S \left(\frac{n_0^2}{36 \pi}\right)^{1/3}~~~\mathrm{and}~~~~~~
\lambda = \alpha_C \left(\frac{n_0}{384 \pi^2}\right)^{1/3}. \label{Eq:sigma_lambda}
\end{equation}
Here $E_0/n_0$ is the minimum energy per baryon of SQM with $n_0$ being the corresponding baryon number density. The results
are then presented in Table~\ref{table:prop} and Fig.~\ref{Fig:SigmaL}. It is found that the surface tension $\sigma$ increases
with the confinement strength parameter $D$ and decreases with $C$. Note that for positive $C$ the corresponding perturbative
interaction is repulsive and a strangelet becomes less bound, while the one-gluon-exchange interaction obtained with $C<0$ has
the opposite effects. For the parameter set ($C=0.7$ and $\sqrt{D} = 129$ MeV according to Fig.~\ref{Fig:MmaxCD}) constrained
with the observational mass of PSR J0348+0432 ($2.01 \pm 0.04\ M_\odot$)~\cite{Antoniadis2013_Science340-1233232}, we find
$\sigma \approx 2.4$ MeV/fm${}^2$ and $\lambda \approx 5.12$ MeV/fm. For other cases, we expect slightly larger values.
It should be pointed out that the constraint for 2 $M_\odot$ strange stars may be relaxed if there exist two separate families
of compact stars~\cite{Drago2014_PRD89-043014}, in which case larger $\sigma$ and $\lambda$ may be expected. Meanwhile, as
indicated in Fig.~\ref{Fig:Surf_ni}, a larger $n_0$ corresponds to more drastic variations of densities and potentials on the
surface of a strangelet, which suggests some correlations of $\sigma$ and $\lambda$ with $n_0$. Indeed, this is observed in
Fig.~\ref{Fig:SigmaL}, where $\sigma$ and $\lambda$ increase with $n_0$ and can be approximated with $\sigma \approx 14.3 n_0
+ 1.3$ and $\lambda \approx 36.6 n_0+1.3$ with the units corresponding to those in Table~\ref{table:prop}.

\section{\label{sec:con}Summary}

We discuss the significant implications of the interface effects on the properties of SQM and the related physical processes.
For absolutely stable SQM, it is found that the interface effects destabilize a strangelet substantially for a small baryon
number. If we adopt a small enough surface tension, there may exist strangelets at certain baryon numbers that are more
stable than others, which will lead to many interesting consequences related to SQM objects. For unstable SQM, it was shown
that adopting different surface tension values will results in very different structures for hybrid stars~\cite{Xia2019}.
To investigate the interface effects, we adopt mean-field approximation for equivparticle model, where the strong interactions
are included with density-dependent quark masses. The properties of spherically symmetric strangelets are obtained adopting
various parameter sets. By fitting the energy per baryon of strangelets to a liquid-drop type formula~\cite{Oertel2008_PRD77-074015},
we estimate the surface tension $\sigma$ and curvature term $\lambda$ of SQM. The parameter dependence on the surface tension
and curvature term are examined, where the obtained $\sigma$ and $\lambda$ increase monotonically with the density of SQM at
zero external pressure. For a parameter set constrained according to the 2$M_\odot$ strange star, the surface tension is
expected to be $\sim$2.4 MeV/fm${}^2$, while it is larger for other cases.

\section{Acknowledgement}
This work was supported by National Natural Science Foundation of China (Grant Nos.~11705163, 11875052, 11621131001,
11575201, 11575190, and 11525524), the Physics Research and Development Program of Zhengzhou University (Grant No.~32410017),
and the U.S. National Science Foundation (Grant No. PHY 1608959). The computation for this work was supported by the
HPC Cluster of SKLTP/ITP-CAS and the Supercomputing Center, CNIC, of the CAS.



%

\end{document}